\documentclass[11pt]{article}

\usepackage[english]{babel}
\usepackage[utf8]{inputenc}
\usepackage[T2A]{fontenc}

\usepackage[top=2cm,bottom=2cm,left=2cm,right=1.5cm,marginparwidth=1.75cm]{geometry}

\usepackage{mathtools}
\usepackage{amsmath}
\usepackage{amsfonts}
\usepackage{amssymb}
\usepackage{bm}
\usepackage[italicdiff]{physics}
\usepackage{derivative}

\usepackage{graphicx}
\usepackage{float}
\usepackage[colorlinks=true, allcolors=blue]{hyperref}

\usepackage{authblk}

\usepackage{indentfirst}
\usepackage{setspace}
\usepackage[x11names]{xcolor}
\usepackage{enumitem}

\title{\textbf{Squeezing for dispersive readout of NV magnetometer}}
\date{}

\author[1,2,3]{A.M. Kozodaev}
\author[1,2,3,4]{S.M. Drofa}
\author[1,2,3]{P.G. Vilyuzhanina}
\author[1,2,3]{A. Chernyavskiy}
\author[2,3]{N.I. Salangin}
\author[4]{A.N. Smolyaninov}
\author[5,6]{S.Ya. Kilin}
\author[1,3,4]{I.S. Cojocaru}
\author[3,4]{S.V. Bolshedvorskii}
\author[3,4]{V.V. Soshenko}
\author[1]{F.Y. Khalili}
\author[1,3,4]{A.V. Akimov}

\affil[1]{Russian Quantum Center, Bolshoy Boulevard 30, building 1, Skolkovo, 121205, Russia}
\affil[2]{Moscow Institute of Physics and Technology, Institutskii pereulok 9, Dolgoprudny, Moscow Region 141701, Russia}
\affil[3]{P.N.Lebedev Physical Institute of the Russian Academy of Science, Leninsky Prospekt 53, Moscow, 119991, Russia}
\affil[4]{LLC Sensor Spin Technologies, 121205 Nobel St. 9, Moscow, Russia}
\affil[5]{National Research Nuclear University “MEPhI”, 31, Kashirskoe Highway, Moscow, 115409 Russia}
\affil[6]{B.I. Stepanov Institute of Physics NASB, 68, Nezavisimosty Ave, Minsk, 220072 Belarus}

\begin{document}

\newcommand{\fk}{}
\newcommand{\sk}{\color{black}}
\newcommand{\rev}{}

\maketitle

\begin{abstract}
\noindent
Nitrogen vacancy centers in diamond {\rev have established themselves} as good sensing element for various type of sensors.
In particular magnetometers based on diamond impurities are quickly developing and are already on the market.
Yet, optical readout in these systems complicates {\rev system design}.
Recently schemes of dispersive readout of nitrogen vacancy spin state using high finesse dielectric cavities for microwave field were proposed, {\fk which do not use the optical readout scheme. However,} only shot noise based estimates were so far {\fk done} for sensitivity of these devices.
Here we provide detailed analysis of various practically relevant {\fk noise and loss sources} for such a system.
Furthermore, we consider the possibility {\fk of using the squeezing quantum state of the probing microwave field and show it allows} to improve the device performance even at room temperature.
\end{abstract}

\section{Introduction}
Today, many research groups work with quantum sensors based on nitrogen vacancy (NV) centers in diamond.
Numerous theoretical and experimental studies have focused on optical readout of NV center states by analyzing {\rev the luminescence signal of optically detected magnetic resonance \cite{Barry_24, Barry_20, Jensen_13}. This technique allowed number of advances with NV centers both in development quantum memory and elements of quantum computing \cite{Maurer_2012, Finsterholzl_2025, Dutt_2007} and for quantum sensing \cite{Wood_2022, Mcguinness_2011, Soshenko_2021, Taylor_2011, Kucsko_2013}. In particular in magnetometry in a range of around $\sim10\text{ pT}\sqrt{\text{Hz}}$ was achieved by many groups at low frequencies \cite{Hahl_2022, Nabeel_2023, Tang_2023}, scanning high resolution probes were realized \cite{Palm_2022, Oshnik_2022, Hochstetter_2021, Rani_2020}, wide range of frequencies was made assessable \cite{Wang_2022}. Besides sub-picotesla sensitivity was also achieved using nuclear spin quantum memory and magnetic field concentrators \cite{Wolf_2015, Xie_2020, Silani_2023}.}

However, recently new works have appeared that read out NV center states by measuring the frequency-dispersive shift in a cavity, which can improve the sensitivity to magnetic fields \cite{Eisenach_2021, Ebel_2021}.
In this approach, {\fk the response of a cavity with the quantum spin system is measured} \cite{Brune_90}.
The quantum system modifies the cavity eigen frequency, and this frequency shift is detected by measuring microwave field reflected from the cavity.
This method is known as dispersive readout.

One of the first theoretical papers on this topic was \cite{Collet1984}, in which the authors presented a general theory of input-output for quantum dissipative systems, which allows connecting the output of a cavity with it input through the internal dynamics of the system.
In \cite{Kohler_2018}, an expression for the cavity transmission coefficient was obtained that goes beyond the rotating wave approximation, and the theory of linear response was described.
The dispersive readout method used in a number of works to create quantum magnetic field sensors based on NV centres or other compounds.
For example, in \cite{wilcox_2021} the authors used $Cr^{3+}$ defects in sapphire crystal ($Al_2O_3$) and developed a device with a sensitivity of 9.7 $pT/\sqrt{Hz}$.
Paper \cite{Ebel_2021} demonstrated that dispersive readout of NV in a cavity can achieve a sensitivity beyond the shot noise limit of optical readout.
The work \cite{Eisenach_2021} investigated the possibility of an NV centre magnetic sensor and suggested that it could reach a sensitivity of 3 $pT/\sqrt{Hz}$.
A key feature of these sensors is that they have much lower noise and readout errors because they do not use optical interaction with the quantum system.

However, in dispersive readout, the sensitivity is fundamentally limited by the quantum noise of the microwave field.
According to the Heisenberg uncertainty principle, the fluctuations of the amplitude and phase quadratures satisfy the uncertainty relation $\Delta a^c\Delta a^s \ge \frac 12$. In the vacuum and coherent quantum states,  $\Delta a^c\Delta a^s = \frac 12$. This gives the shot-noise limit (SNL) for phase sensitivity measurements \cite{Caves_1981, Caves_1982}.
Squeezed states offer way to beat this limit.
In squeezing, quantum noise is redistributed among the quadratures: noise in one quadrature drops below the vacuum level, while noise in the perpendicular quadrature increases \cite{Caves_1981, Castellanos_2008, Macklin_2015, 20a1FrAgKhCh, Khalili_2025}.
If we choose the squeezing angle so that the squeezed quadrature aligns with the one that carries the frequency-shift signal, we can improve the signal-to-noise ratio without raising the probe power.
The generation of squeezed microwave fields already has been demonstrated experimentally using Josephson parametric amplifiers \cite{Danner_2025}.
This makes their use in solid-state magnetometers practically feasible.

Despite progress in dispersive readout, full theoretical analysis of how squeezed states affect the sensitivity of NV center magnetometers is still missing.
In particular, it remains unclear how squeezing interacts with thermal spin noise, cavity losses, and the finite efficiency of homodyne detection.

In this paper, we develop a theory of dispersive readout magnetometer based on an ensemble of NV centers in a microwave cavity \cite{Soshenko_2020, Kapitanova_2018}, which takes into account quantum properties of the scanning microwave (MW) field (including compressed states) and all the main noise sources limiting sensitivity.
The purpose of this work is to derive an analytical expression for the spectral density of noise and sensitivity during dispersive readout method and use of squeezed states to suppress noise in the input electromagnetic MW field.


As a result, we obtain an analytical expression for the sensitivity of a dispersive magnetometer that includes input noise, cavity noise and spin noise.
We also show how squeezed states can improve the sensitivity.


\begin{figure}[h]
    \centering
    \includegraphics[width=\columnwidth]{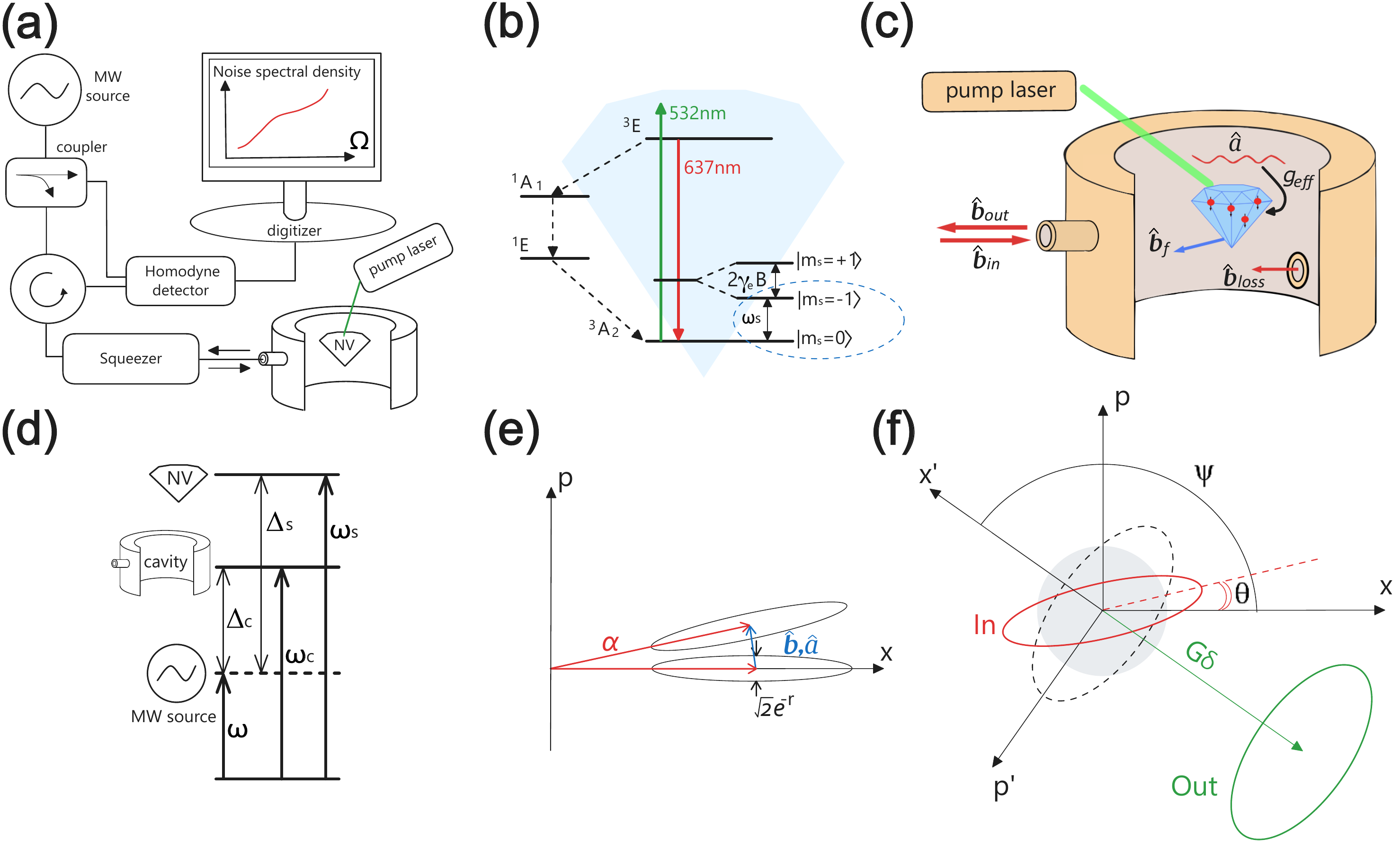}  
    \caption{\fk
    a) Scheme of the experiment.  The external MW field passes through the squeeze element. The signal reflected from the cavity falls on the homodyne detector. As a result, we get the noise spectral density.
    b) NV center in diamond principal level scheme. The two-level
    system selected for magnetometry encircled by blue dot-dashed line;
    c) Cavity scheme. $\hat b_{in}, \hat b_{out}$ - input/output quantum noise of MW field; $\hat b_{loss}$-thermal noise of the cavity; $\hat b_{f}$- NV spin noise; $g_{eff}$-coupling coefficient; $a$-cavity mode operator
    d) Principal scheme of the frequencies used;
    e) Idea of separation of classical offset field and quantum signal, which is assumed much smaller then classical;
    f) Parameters of the quantum part only. Covariance ellipses of the output field and the direction of homodyne detection $\psi$. Gray circle - initial coherent state, red - input squeezed state, dotted line - output field in the absence of a signal $\delta$ and green - output state with a signal where the center shift is proportional to the amplitude of the magnetic field.}\label{fig:scheme1}
\end{figure}

\section{Evolution of probe field}\label{section_Evolution_of_probe_field}
\subsection{Hamiltonian and model}
The scheme of the modeled experiment is shown in Fig.\,\ref{fig:scheme1}a.
We consider a quantum system consisting of an ensemble of nitrogen-vacancy (NV) centers in
diamond placed inside a microwave cavity.
NV centers in diamond have total spin of 1 forming triplet in the ground state, in which component on the NV axis with $|m_s=\pm1\rangle$ and $|m_s=0\rangle$ have splitting of about 2.87 GHz (see Figure \ref{fig:scheme1}b).
Under green laser radiation the system tends to polarize into $|m_s=0\rangle$ state thus creating non-equilibrium population of states.
The $|m_s=\pm1\rangle$  split in the presence {\fk of magnetic field thus allows} magnetometry.
In order to consider coupling of NV centers to the dielectric cavity we approximate system of $|m_s=0\rangle$
and $|m_s= -1\rangle$ as an effective two-level system (TLS) with resonance frequency $\omega_{sys}$.
Laser radiation pumping is assumed to be done before microwave experiment.

The cavity is characterized by the eigenfrequency $\omega_{cav}$ and creation and annihilation operators $a,a^+$ of photon mode, {\fk see} Fig.\,\ref{fig:scheme1}c,d.
The interaction between spins and cavity field is dipole and is described by the coupling constant $g_1$.
In the laboratory frame, total Hamiltonian of the system is:
\begin{equation}\label{Ham1}
    H = \frac{\omega_{sys}}{2}\sigma_z + \omega_{cav}a^+a + g_1(\sigma_+ + \sigma_-)(a+a^+),
\end{equation}
where  $\sigma_z,\sigma_\pm$ are the Pauli operators of the TLS.
The terms $a\sigma_-$ and $a^+\sigma_+ $ oscillate rapidly at frequencies $\omega_{cav} + \omega_{sys}$.
In typical experiments, {\fk contribution of these terms to system dynamics is very small.}
We use the rotating-wave approximation (RWA) and neglect them:
\begin{equation}\label{Ham2}
    H = \frac{\omega_{sys}}{2}\sigma_z + \omega_{cav}a^+a + g_1(a\sigma_+ + a^+\sigma_-),
\end{equation}

To analyze dispersive readout, we go to rotating frame at the frequency $\omega$ of the external microwave field with unitary transformation $U=e^{-i\omega t(\frac{\sigma_z}{2} + a^+a)}$ and detunings $\Delta_{s}=\omega_{sys}-\omega,~\Delta_{c}=\omega_{cav}-\omega$ as shown in the Figure \ref{fig:scheme1}d:
\begin{equation}\label{Ham3}
    H = \frac{\Delta_{s}}{2}\sigma_z + \Delta_{c}a^+a + g_1(a\sigma_+ + a^+\sigma_-),
\end{equation}
We now generalize the above description to an ensemble of $N_q$ identical NV centers.
We assume that all centers have the same detuning and the same coupling constant $g_1$.
Interactions between different NV centers are neglected which is a good approximation at low defect concentrations.
The Hamiltonian \eqref{Ham3} is generalized as follows:
\begin{equation}\label{Ham4}
    H = \Delta_{s}J_z + \Delta_{c}a^+a + g_1(aJ_+ + a^+J_-),
\end{equation}
Here we introduced the collective operators $J_z,J_\pm$ which satisfy the standard commutation relations:
\begin{equation}
    J_-=\sum_j^{N_{q}}\sigma_-^{(j)}, ~~
    J_+=\sum_j^{N_{q}}\sigma_+^{(j)}, ~~
    J_z=\frac12 \sum_j^{N_{q}}\sigma_z^{(j)},
\end{equation}
For small spin excitations, we can linearize using the Holstein–Primakoff transformation \cite{Holstein_1940, Hammerer_2008}.
We take all NV centers to be initially polarized, so that $\langle J_z \rangle \approx -\frac{N_{q}}{2}$.
In other words, most spins are in the ground state, and the number of collective excitations is small: $\langle d^+d\rangle \ll N_{q}$.
Under this condition, $J\pm$ can be replaced by bosonic operators $d,d^+$ with  $[d,d^+]=1$:
\begin{equation}\label{boson_operators}
    d = \frac{J_-}{\sqrt{2\langle J_-\rangle}} \approx \frac{J_-}{\sqrt{N_{q}}},  ~~~d^+\approx \frac{J_+}{\sqrt{N_{q}}}, ~~~J_z\approx -I\frac{N_{q}}{2}+d^+d
\end{equation}
{\rev The main condition for validity of the  Holstein–Primakoff approximation in the weak excitation of the spins ensemble.
For the parameters that we use for our estimates, see Tab.\ref{tab:sim_params}, the mean value of the spin excitations $N \times g_{eff}^2/\Delta_s^2$ is equal to $\approx 4 \times 10^{12}$, or about 4\% of the total number of spins.}
Inserting operators \eqref{boson_operators} into the Hamiltonian \eqref{Ham4} and omitting the constant term $-\frac{N_q}{2}\Delta_{sys}$, as it only shifts the energy, we obtain:
\begin{equation}\label{Ham5}
    H = \Delta_{s}d^+d + \Delta_{c}a^+a + g_{eff}(ad^+ + a^+d)
\end{equation}
Here we defined an effective coupling constant $g_{eff}=g_1\sqrt{N_q}$.

\subsection{Langieven equations}

The Hamiltonian \eqref{Ham5} describes the dynamics of a closed system.
But in practice the system open and cavity is connected to external waveguides for microwave input and output and the spins experience relaxation and dephasing.
To include dissipation and the related quantum noise, we use the quantum Langevin equations.
Within Markov approximation, equations for the cavity mode $a$ and the HP operator $d$ are:
\begin{equation}\label{Langeiven_eq}
    \begin{cases}
        \dot{a}=-(i\Delta_c+\frac k2)a - ig_{eff}d +\sqrt{k_1}b_{in, 1} + \sqrt{k_2}b_{in, 2} \\
        \dot{d} = -(i\Delta_s+\frac{\gamma}{2})d - ig_{eff}a + \sqrt{\gamma} b_f
    \end{cases}
\end{equation}
Here $k=k_1+k_2$ - cavity linewidth; $k_1$ - loss rate through the input; $k_2$ - losses inside the cavity; $b_{in, 1}$ - input electromagnetic MW field; $b_{in, 2}$ - vacuum or thermal noise {\fk originating from the losses} inside the cavity; $\gamma$ -  coherence loss rate, which is determined by {\rev $\gamma = 1/T_2^*$}; $b_f$ - spin noise operator associated with relaxation of the quantum system.
{\rev All three noise operators $b_{in,1}$ , $b_{in,2}$, $b_f$ have zero mean values.
Their spectral densities, in principle, can be frequency-dependent.
However, within the narrow frequency band $k\ll\omega_{cav}$ of the high-Q cavity, these spectral dependencies can be neglected \cite{Aspelmeyer_RMP_86_1391_2014, 12a1DaKh}.
As a result, the noises can be considered as a delta-correlated ones with the correlation functions proportional to some effective temperature $T=\hbar\omega_{cav}n_T/\kappa_B$.
For example, for the spin noise at temperature $T$ the correlation function is the same as for harmonic oscillator:}
\begin{equation}
    <b_f(t)b_f^+(t')> = (n_T + 1)\delta(t-t'),~~~ <b_f^+(t)b_f(t')> = n_T\delta(t-t').
\end{equation}

\subsection{Linearization}

When input MW field is strong and coherent, we can split operators in \eqref{Langeiven_eq} into classical mean values plus small quantum fluctuations (Fig.\ref{fig:scheme1}e):
\begin{equation}\label{Liner}
    a = \alpha + \hat{a}, ~~~
    b_{in, 1} = \beta_{in} + \hat{b}_{in}, ~~~
    b_{in, 2} = \hat{b}_{loss}, ~~~
    d = \xi + \hat{d}, ~~~
    b_f = \hat b_f
\end{equation}
Here $(\alpha, \beta_{in, 1}, \xi)$ are classical amplitudes, and $(\hat{a}, \hat{b}_{in},b_{in, 2}, \hat{b}_{loss},\hat{d})$ are quantum fluctuation operators with zero mean.
Note that spin noise $F_d$ and $b_{in, 2}$ have not classical part.
We also write the NV detuning as $\Delta_{sys}=\Delta_{sys}^0+\delta$ where $\Delta_{sys}^0$ is constant and $\delta$ is a readout signal.
{\sk The main quantum noises in the system and idea of separation of classical offset field and quantum signal are shown in the Figure \ref{fig:scheme1}c,e.}

Substituting \eqref{Liner} into the \eqref{Langeiven_eq} system and averaging over the quantum state we obtain a system of equations for classical amplitudes in the stationary regime ($\dot\alpha =0$, $\dot\xi =0$):
\begin{equation}\label{Lang_classic}
    \begin{cases}
        0=(-i\Delta_c-\frac k2)\alpha - ig_{eff}\xi +\sqrt{k_1}\beta_{in}\\
        0 = (-i\Delta_s^0-\frac{\gamma}{2})\xi - ig_{eff}\alpha
    \end{cases}
\end{equation}
The  solution:
\begin{equation}\label{eq_HP_amplitude}
    \alpha = \frac{\sqrt{k_1}\beta_{in, 1}}{i\Delta_c + \frac k 2 + \frac{g_{eff}^2}{i\Delta_s^0+\frac \gamma 2}}, ~~~
    \xi = \frac{-ig_{eff}\alpha}{i\Delta_s^0+\frac \gamma 2}
\end{equation}
Subtracting the classical equations \eqref{Lang_classic} from the full Langevin equations \eqref{Langeiven_eq} we obtain linearized system for quantum fluctuation operators:

\begin{equation}\label{Lang_fluctuations}
\begin{cases}
    \dot{\hat{a}}=-(i\Delta_c+\frac k2)\hat a - ig_{eff}\hat d +\sqrt{k_1}\hat b_{in} + \sqrt{k_2}\hat b_{loss}\\

    \dot{\hat d} = -(i\Delta_s^0+\frac \gamma 2)\hat d - i\xi\delta - ig_{eff}\hat a + \sqrt{\gamma} \hat b_f
\end{cases}
\end{equation}

\subsection{Output field}

The system \eqref{Lang_fluctuations} is linear in the fluctuation operators.
This allows us to solve it in the frequency domain.
Using the Fourier transform, we derive the equation for the cavity mode $a(\Omega)$ in frequency space, see Appendix \ref{Appendix_output_field} for details.
For compactness we introduce the propagators for cavity and HP operators:
\begin{equation}
    P_a(\Omega) = -i\Omega +i\Delta_c+\frac k2,  ~~~P_d(\Omega) = -i\Omega +i\Delta_s^0 + \frac \gamma 2.
\end{equation}
When there is no coupling between the NV centers and the cavity ($g_{eff}=0$), the caity's spectral response to an external perturbation is given by $1/P_a(\Omega)$.
This function peaks when the analysis frequency equals the cavity detuning from the pump $\Omega=\Delta_c$.
The resonance width is {\rev $k$}.
Similarly,  $1/P_d(\Omega)$ describes the response of the spin system, with a resonance at $\Omega=\Delta_{sys}^0$ and width $\fk\gamma$.

Using the input-output relation $\hat b_{out}(\Omega) = \sqrt{k_1}\hat a(\Omega) - \hat b_{in}(\Omega)$ we get the final expression for the output field fluctuations $\hat b_{out}(\Omega)$:
\begin{equation}\label{b_out}
    \hat b_{out}(\Omega) = G(\Omega)\delta(\Omega) + K_{in}(\Omega)\hat b_{in}(\Omega) + K_{loss}(\Omega)\hat b_{loss}(\Omega) + K_{f}(\Omega)\hat b_f(\Omega)
\end{equation}
The functions $G(\Omega), K_{in}(\Omega), K_{loss}(\Omega), K_f(\Omega)$ are defined in Appendix \ref{Appendix_output_field}.

Equation \eqref{b_out} is the key result of this section.
It fully describes the output field in the frequency domain and includes four terms.
The signal term $G(\Omega)\delta(\Omega)$ is proportional to the magnetic field fluctuations $\delta(\Omega)$.
The transfer function $G(\Omega)$ determines how sensitive the system is to the magnetic signal.
The input noise $K_{in}(\Omega)\hat b_{in}(\Omega)$ comes from quantum fluctuations of the microwave field sent into the cavity.
The loss noise $K_{loss}(\Omega)\hat b_{loss}(\Omega)$ rises from vacuum or thermal noise entering through loss channels.
This noise is unavoidable and defines the best possible sensitivity.
The spin noise $K_f(\Omega)\hat b_f(\Omega)$ comes from thermal and quantum fluctuations of the spins.

Equation \eqref{b_out} also shows that noise can be suppressed with squeezing.
If we prepare the input field $\hat b_{in}(\Omega)$ so that it fluctuations are small in a given quadrature and if the homodyne detector measure that quadrature, then the input noise term $K_{in}(\Omega)\hat b_{in}(\Omega)$ can be strongly reduced.

\section{Noise spectral density}\label{section_Noise_spectral_density}
\subsection{Homodyne readout}

Equation \eqref{b_out} describes the output field as a complex operator in the frequency domain.
However, in a homodyne detection experiment, we do not measure the operator itself. Instead, we measure its projection onto a particular quadrature, which is set by the local oscillator phase.
To analyze the noise properties and to include squeezed states, we need to switch from complex amplitudes to quadrature components.
We define the amplitude (cosine) and phase (sine) quadrature operators for the quantum noise operators $b(\Omega)$ as follows \cite{Caves1985, Schumaker1985}:

\begin{equation}\label{b_c_b_s}
    b^c(\Omega) = \frac{b(\Omega) + b^+(-\Omega)}{\sqrt2}, ~~~b^s(\Omega) = \frac{b(\Omega) - b^+(-\Omega)}{i\sqrt2}
\end{equation}
Homodyne detection makes it possible to measure the projection of the output field $\hat b_{out}$ onto an arbitrary quadrature determined by the phase of the local oscillator $\psi$ (Figure \ref{fig:scheme1}f).
The output signal of the homodyne detector is proportional to the operator:
\begin{equation}\label{eq_Homodine_detect}
    \hat b_{out}^{\psi}(\Omega) = \hat b_{out}^c(\Omega)\cos{\psi} + \hat b_{out}^s(\Omega)\sin{\psi}
\end{equation}
This equation gives the field projection onto a quadrature rotated by $\psi$ from the amplitude quadrature.
By tuning the local oscillator phase $\psi$ we can measure the amplitude quadrature, the phase quadrature or anything in between.

To write \eqref{eq_Homodine_detect} in terms of the transfer functions $G(\Omega)$ and $K_i(\Omega)$ we introduced the quadrature components $G^c(\Omega), G^s(\Omega)$ and $K_i^c(\Omega), K_i^s(\Omega)$, see Appendix \ref{Appendix_homodyne_detection} for details.
Expression \eqref{eq_Homodine_detect} then becomes:
\begin{equation}\label{eq_Homodine_detect_2}
    \hat b_{out}^{\psi}(\Omega) = G_{\psi}(\Omega)\delta(\Omega) + \sum_{i = \{in, loss, f \}}(A_ib^c_i + B_ib^s_i),
\end{equation}
where
\begin{equation}
    G_\psi(\Omega) = G^c(\Omega)\cos{\psi} + G^s(\Omega)\sin{\psi}
\end{equation}
\begin{equation}
    A_i = K^c_i(\Omega)\cos{\psi} + K^s_i(\Omega)\sin{\psi}
\end{equation}
\begin{equation}
    B_i = - (K^s_i(\Omega)\cos{\psi} - K^c_i(\Omega)\sin{\psi})
\end{equation}
From \eqref{eq_Homodine_detect_2} we see that the homodyne signal consists of the useful signal $G_{\psi}(\Omega)\delta(\Omega)$ proportional to magnetic fluctuations and noise from quadratures of each noise source, with weights $A_i$ and $B_i$.
Since these coefficients depend on $\psi$, we can tune the local oscillator phase to minimize the noise we do not want.

\subsection{Spectral density}

To calculate noise spectral density, it is necessary to know the correlation functions of the operators $\hat b^c_i$ and $\hat b^s_i$.
For all three noise sources the correlation functions can be written in a uniform form, as shown in Appendix \ref{Appendix_spectral_densities}.
For the i-th source (i=in,loss,f) we have:
\begin{subequations}\label{eq_correlate_function}
\begin{equation}\label{eq_correlate_function_1}
    \langle b_i^c(\Omega)b_i^c(-\Omega') \rangle
    = 2\pi S^{cc}_i\delta(\Omega-\Omega'),
\end{equation}
\begin{equation}\label{eq_correlate_function_2}
    \langle b_i^s(\Omega)b_i^s(-\Omega') \rangle
    = 2\pi S^{ss}_i\delta(\Omega-\Omega'),
\end{equation}
\begin{equation}\label{eq_correlate_function_3}
    \frac 12\langle \{b_i^c(\Omega)b_i^s(-\Omega') \} \rangle = 2\pi S^{cs}_i\delta(\Omega-\Omega'),
\end{equation}
\end{subequations}
The spectral densities $S^{cc}_{in},S^{ss}_{in}, S^{cs}_{in}$ depend on the squeeze angle $\theta$ and the squeeze parameter r and $S^{cs}_{loss}=S^{cs}_{f} = 0$.
Using the expression for the homodyne signal \eqref{eq_Homodine_detect_2} and the spectral densities \eqref{eq_correlate_function_1}, \eqref{eq_correlate_function_2}, \eqref{eq_correlate_function_3} we obtain the spectral noise density reduced to the measured signal $\delta(\Omega)$:
\begin{equation}\label{eq_spectral_dens}
    S_{\delta} = \frac{1}{|G_\psi|^2}\sum_{i = \{in, loss, f \}}(K^{cc}_iS^{cc}_i + K^{ss}_iS^{ss}_i + 2K^{cs}_iS^{cs}_i )
\end{equation}
Here, coefficients $K^{cc}_i, K^{ss}_i, K^{cs}_i$ are determined through the transfer functions $A_i$ and $B_i$:
\begin{equation}\label{K_i}
    K_i^{cc} = |A_i|^2 = |K_i^c\cos\psi + K_i^s\sin\psi|^2
\end{equation}
\begin{equation}
    K_i^{ss} =|B_i|^2 = |K_i^s\cos\psi - K_i^c\sin\psi|^2
\end{equation}
\begin{equation}
    K_i^{cs} = Re\{ A_iB_i^*\} = Re\{(K^c\cos\psi + K^s\sin\psi)(K^c\sin\psi - K^s\cos\psi)^*  \}
\end{equation}
Equation \eqref{eq_spectral_dens} provides a complete expression for noise spectral density at homodyne output, normalized to magnetic signal $\delta$. It accounts for the input field (including arbitrary squeezing), cavity losses, spin noise and the local oscillator phase $\psi$.

\subsection{Optimization by squeeze angle}

The resulting noise spectral density consists of the contribution of the input field, cavity noise, and spin noise:
\begin{equation}
    S_{\delta} = S_{\delta, in} + S_{\delta, loss} + S_{\delta, f}
\end{equation}
The noise spectral density  $S_{\delta, in}(r, \theta)$ depends on the squeezing parameters ${\rev r}$ {\rev and squeezing angle} $\theta$, which we can control in experiment.
We want to find the angle $\theta_{opt}$ that minimizes the input noise contribution.
This corresponds to the best use of squeezing to suppress quantum fluctuations in the measured quadrature.
Inserting the correlation functions of the squeezed state gives:
\begin{equation}\label{eq_s_delta_in}
    S_{\delta, in} =\frac{n_{T}+\frac12}{|G_\psi|^2} \left[(K^{cc}_{in} + K^{ss}_{in})ch~2r + (K^{cc}_{in} - K^{ss}_{in})sh~2r\cos{2\theta} + 2K^{cs}_{in} sh~2r\sin{2\theta}\right]
\end{equation}
To find the minimum, we differentiate \eqref{eq_s_delta_in} by $\theta$:
\begin{equation}
    \frac{\partial S_{\delta, in}}{\partial\theta} =
    \frac{2n_{T}+1}{|G_\psi|^2} ((K^{ss}_{in} - K^{cc}_{in})\sin{2\theta} + 2K^{cs}_{in}\cos{2\theta})sh(2r) = 0
\end{equation}

We choose the optimal squeezing angle $\theta_{opt}$ as the angle that minimizes the noise spectral density at the resonance ($\Omega=0$).
It give condition for the optimal squeezing angle:
\begin{subequations}
\begin{gather}
  \sin{2\theta_{opt}} = -\frac{2K^{cs}_{in}(\Omega=0)}{          \sqrt{[K^{cc}_{in}(\Omega=0) - K^{ss}_{in}(\Omega=0)]^2
        + [2K^{cs}_{in}(\Omega=0)]^2}
      }  \,, \\
  \cos{2\theta_{opt}} =
    -\frac{K^{cc}_{in}(\Omega=0) - K^{ss}_{in}(\Omega=0)}{       \sqrt{[K^{cc}_{in}(\Omega=0) - K^{ss}_{in}(\Omega=0)]^2
      + [2K^{cs}_{in}(\Omega=0)]^2} \,,
  }
\end{gather}
\end{subequations}
Inserting $\theta_{opt}$ into \eqref{eq_s_delta_in} we obtain the minimum value of the input noise contribution $S_{\delta, in}(\theta_{opt})$. The optimal squeezing angle $\theta_{opt}$ orients the squeezed state so that its squeezed axis coincides with the quadrature that contributes most to the total spectral density.

{\sk Similarly, we use the equation $\frac{\partial S_{\delta}}{\partial\psi}=0$ to find the optimal homodyne angle $\psi_{opt}$ that provides the best sensitivity.}
Since $S_{\delta}$ depends on $\psi$ in all it terms, we minimize the full spectral density.

\begin{figure}[ht]
\centering
\includegraphics[width=1.\textwidth]{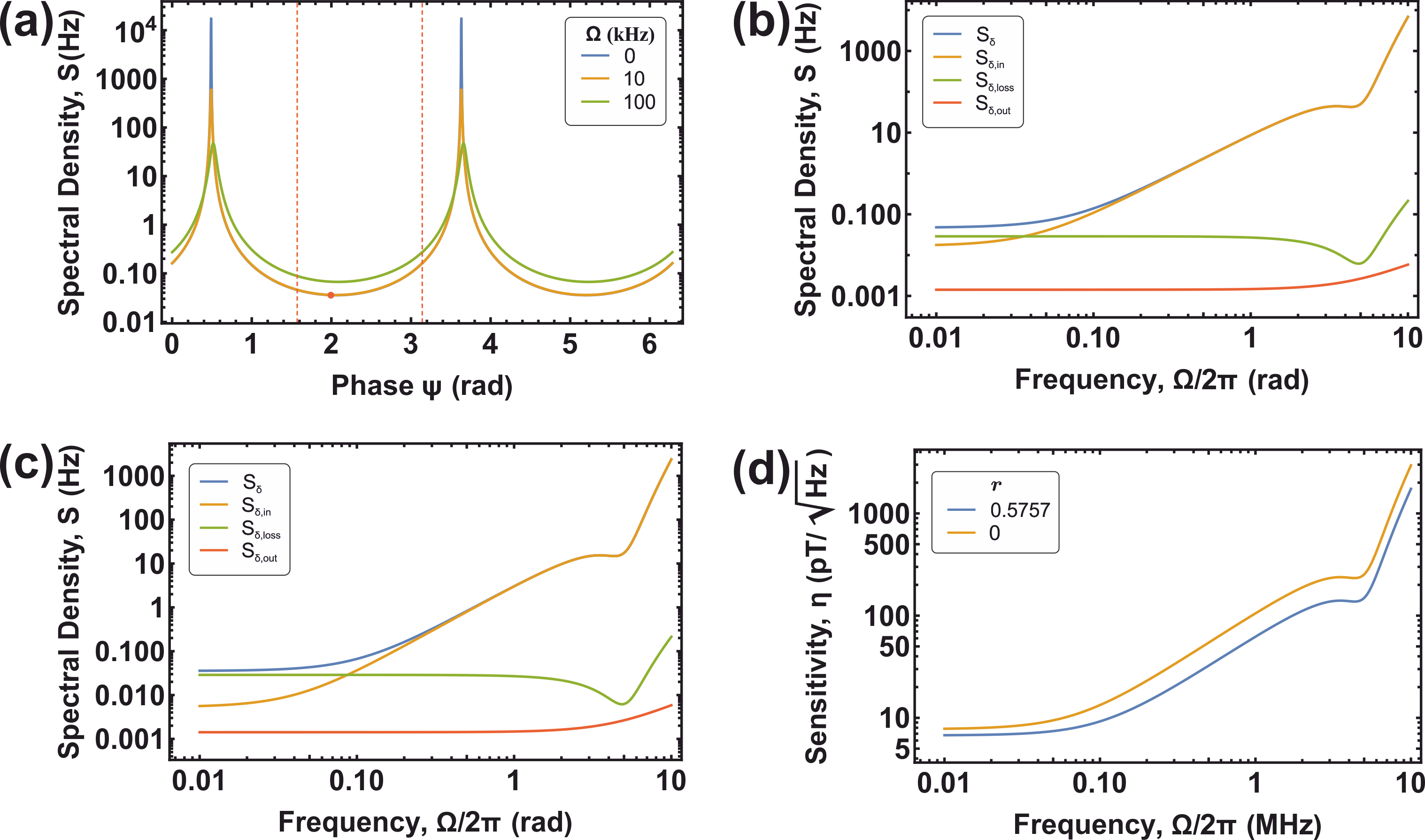}
\caption{a) Dependence of spectral density on $\psi$ for different $\Omega$ at $\theta_{opt}\approx 1.35$rad at room temperature (T=293K). The red marker on the picture indicates the minimum at $\Omega=0$. {\rev The simulation parameters are summarized in Table \ref{tab:sim_params}:} $\Delta_c=0, \Delta_s=2\pi \times5$MHz, $Q=2\times 10^4$, $k_1=0.8k$, the number of photons in the cavity $N=6.2 \times 10^{15}$, {\rev the number of NV centers $N_q=9 \times 10^{14}$, ensemble coupling coefficient $g_{eff} = 2\pi \times 0.12~ MHz$ (see Appendix \ref{Appendix_coupling_coefficient}), squeeze parameter $r=0$ (no squeeze)};
b,c) Contribution of individual components to the spectral density, b - no squeeze, c - squeeze of 5 dB;
d) Sensitivity with 5 dB squeezing (blue) and without squeezing (orange).}\label{fig:results1}
\end{figure}

In this approach both $\theta_{opt}$ and $\psi_{opt}$ are fixed and do not depend on $\Omega$.
These angles can be set once in the experiment.
{\sk The figure \ref{fig:scheme1}f) shows the covariance ellipses of the output field and the direction of homodyne detection. }
The resulting minimum is shown in Fig. \ref{fig:results1}a for the resonance frequency $\Omega=0$.
The red marker indicates the point where the spectral density reaches its minimum at resonance.
The total noise spectral density, minimized over $\theta$ and $\psi$, is then:

\begin{equation}\label{eq_final_spectral_density}
    \min_{\theta, \psi} S_{\delta} =S_{\delta}(\theta_{opt},\psi_{opt}) = S_{\delta, in}(\theta_{opt},\psi_{opt}) + S_{\delta, loss}(\psi_{opt}) + S_{\delta, f}(\psi_{opt})
\end{equation}
This expression is the key result of the work.
It provides a complete analytical description of the spectral noise density reduced to the $\delta(\Omega)$ signal.
The contribution of each of the terms to the total spectral density is shown in the Fig.\ref{fig:results1}b,c.
As can be seen from the graph, at low $\Omega$ the spectral density is limited not by the noise of the input field $\hat b_{in}$, but by the noise of the interaction of the field with the cavity $\hat b_{loss}$.

\section{Sensitivity}\label{section_sensitivity}
The sensitivity $\eta$ of the magnetometer is defined as the minimum magnetic field detectable in 1 second.
In the frequency domain, the sensitivity is expressed through the noise spectral density $S_{\delta}(\Omega)$ and the conversion factor that relates magnetic field to frequency detuning $\delta$.
The conversion is set by the electron gyromagnetic ratio  ${\rev \gamma_e} = 28~GHz/T$: $\delta ={\rev \gamma_e}\Delta B $.
Therefore, the noise spectral density referred to the magnetic field is $S_B(\Omega) = S_{\delta}(\Omega)/{\rev \gamma_e^2}$.
The sensitivity (in $T/\sqrt{Hz}$) is then the square root:
\begin{equation}\label{eq_final_sensitivity}
    \eta=\sqrt{S_B(\Omega)} = \frac{\sqrt{S_{\delta}(\Omega)}}{{\rev \gamma_e}}
\end{equation}
With this expression, we can see how various parameters influence the magnetometer sensitivity and find the best settings for the experiment. The sensitivity at room temperature is shown in Fig.\ref{fig:results1}d.

\begin{figure}[H]
\centering
\includegraphics[width=1.0\textwidth]{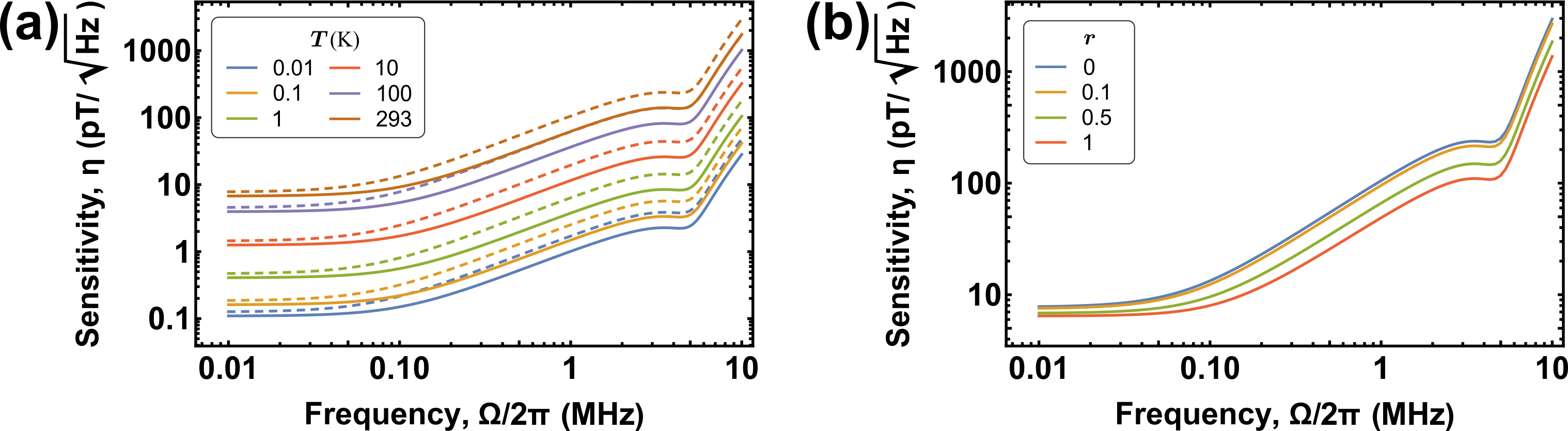}
\caption{a) Sensitivity versus temperature. Solid curves correspond to 5 dB squeezing (${\rev r}$=0.5757), dashed curves to no squeezing. Data are shown for T = 0.01, 0.1, 1, 10, 100, and 293 K.; b) Room-temperature (T=293 K) sensitivity for various squeezing parameters r. {\rev Other modeling parameters correspond to the fig.\ref{fig:results1} and summarized in Table \ref{tab:sim_params}.}}\label{fig:results2}
\end{figure}

Figure \ref{fig:results2}a presents the sensitivity $\eta(\Omega)$ as a function of the analysis frequency $\Omega$ for various temperatures, ranging from very low (T=0.01 K) to room temperature (T=293 K).
Solid lines correspond to 5 dB squeezing (${\rev r}$=0.5757), while dashed lines correspond to no squeezing (${\rev r}$=0).
We note that at all temperatures, squeezing gives the largest improvement at high frequencies, where quantum noise dominates.
At low frequencies, the squeezing effect is minimal, because here the main limitation comes from cavity loss noise, which is not affected by squeezing.
These results show that squeezed states are most effective under cryogenic conditions.
However, even at room temperature squeezing remains a useful tool for improving the sensitivity of the magnetometer.

Figure \ref{fig:results2}b presents the sensitivity $\eta(\Omega)$ at room temperature (T=293 K) versus frequency
$\Omega$ for several squeezing parameters ${\rev r}$.
The curves use fixed angles $\theta = \theta_{opt}$ and $\psi = \psi_{opt}$ chosen for the case $\Omega=0$.
We see that stronger squeezing improves the sensitivity across all frequencies.
The improvement is largest at high frequencies, where input quantum noise $S_{\delta,in}$ is the main contributor.
At low frequencies, squeezing helps less because cavity loss noise $S_{\delta,loss}$ is dominant and is not affected by squeezing.

It is worth noting that even moderate squeezing ${\rev r}$=0.5 that corresponding to about 4.3 dB, gives a noticeable improvement in sensitivity.
This makes squeezed states promising for practical applications, even at room temperature.

{\rev It is interesting to compare calculated sensitivity with often used spin projections noise \cite{Barry_20, Eisenach_2021}:
\begin{equation}
  \eta _{multi}^{ensemble} \approx \sqrt{3}\frac{\hbar}{{{g}_{L}}{{\mu }_{B}}}
    \frac{1}{\sqrt{{{N}_{q}}T_{2}^{*}}}\cdot\sqrt{1+\frac{1}{{{C}^{2}}{{n}_{avg}}}}
  \approx 1.65~fT/\sqrt{Hz} \,,
\end{equation}
where ${{g}_{L}}$ is the NV electronic g-factor, ${{\mu }_{B}}$ is the Bohr magneton, $\hbar $ is Planck’s constant, $C=1$ is the readout contrast and ${{n}_{avg}}$ is the average number of MW photons collected per NV- per measurement. This value is significantly lower than the sensitivity predicted by our full model, since it only takes into account the fundamental quantum noise rather than much stronger noises originating from cavity losses, thermal spin noise and quantum noise of the microwave field.}

\section{Discussion}
In this work, we developed theory of a dispersive magnetometer based on an ensemble of NV centers placed in a microwave cavity, including squeezed input states.
The main result is an analytical expression for the noise spectral density \eqref{eq_final_spectral_density} and the sensitivity \eqref{eq_final_sensitivity}.
These expressions include in unified way all main noise sources, quantum noise of the input field, cavity losses and thermal noise of the spin system.


Our main conclusion is that input squeezing can reduce quantum noise if we properly optimize both the squeezing angle $\theta$ and the homodyne phase $\psi$.
But we should keep in mind that the best squeezing angle changes with the frequency $\Omega$.
This makes the method more difficult to apply in broadband experiments.
{\rev A possible solution is to implement a frequency dependent squeezing angle.
Technology of generation of the frequency-dependent squeezing using additional so-called filter cavities was proposed in \cite{02a1KiLeMaThVy} and is used in the laser gravitation-wave detectors LIGO \cite{Tse_PRL_123_231107_2019_short}.}

The analysis of the sensitivity $\eta(\Omega)$ shows that input squeezing does improve the sensitivity.
However, the efficiency of squeezing depends strongly on temperature, frequency range and other experimental parameters.
At room temperature (T=293 K) thermal noise is large and the squeezing effect is partly masked.
Still, even in this case squeezing gives a noticeable improvement in the high-frequency region.
This suggests that squeezed states could be used in practical magnetometers operating at room temperature without the need for cryogenic cooling.
{\rev The developed approach could be used for multy-frequency cavities if vector
magnetometry is desired provided cavity resonances do not overlap.}

There are some limitations in our model that should be considered when interpreting the results or designing future experiments.
The Holstein–Primakoff transformation we used is linearized and only works for small spin excitations.
We also neglected interactions between NV centers.
At high densities, dipole–dipole interactions between spins can lead to extra dephasing and line broadening, making the sensitivity worse.
{\rev The mechanism is that magnetic dipole fields from neighbour NV centers create random local field variations across the ensemble, which accelerate spin decoherence. For the concentration used in this work (n = 1 ppm), the average inter-spin distance is sufficiently large so that dipole–dipole interactions between different NV centers are negligible \cite{Barry_20}. This ensures that our assumption of non-interacting spins remains valid under the chosen experimental conditions.}

In conclusion, our work demonstrates that it is fundamentally possible to overcome the standard quantum limit in NV center magnetometry by using squeezed states.
Use of the squeezed states can also improve sensitivity of a dispersive-readout magnetometer even at room temperature.
The analytical expressions for the sensitivity and the noise spectral density in dispersive readout have been derived.
These results provide researchers with a practical tool for optimizing dispersive readout based magnetometry on NV centers or other similar two-level systems and choosing the best operating regime.
{\sk
\section{Acknowledgments}
	This study was supported by the Ministry of Science and Higher Education of the Russian Federation (agreement/grant No. 075-15-2024-556).
}

\bibliographystyle{ieeetr}
\bibliography{sample}

\newpage
\begin{center}
    \textbf{\Large{Supplementary Materials:}}

    \textbf{\Large{Squeezing for dispersive readout of NV magnetometer}}
\end{center}

\appendix
\section{Calculation of Output Field}\label{Appendix_output_field}
In this section, we show in detail the output of the expression for the output field operator $\hat b_{out}(\Omega)$ from the linearized Langevin equations.
This conclusion is key for obtaining transfer functions and analyzing the noise properties of the system.
Applying the Fourier transform of $X(\Omega) = \int dt e^{i\Omega t} X(t)$ to a system of linearized equations \eqref{Lang_fluctuations}, we obtain a system of algebraic equations:

\begin{equation}
    \begin{cases}
        (-i\Omega +i\Delta_c+\frac k2)\hat a(\Omega) =
        - ig_{eff}\hat d(\Omega) + \sqrt{k_1}\hat b_{in, 1}(\Omega) + \sqrt{k_2}\hat b_{loss}(\Omega)\\
        (-i\Omega +i\Delta_s^0 + \frac \gamma 2)\hat d(\Omega) = -i\xi\delta(\Omega) - ig_{eff}\hat a(\Omega) + \sqrt{\gamma}\hat{b}_f(\Omega)
    \end{cases}
\end{equation}
Here $\Omega$ is the frequency at which we look at the output signal fluctuations.
It is the sideband frequency that comes from modulation by the signal $\delta(t)$ or from quantum noise.
To keep things concise, we define the propagators (susceptibilities) for the cavity and magnon modes:
\begin{equation}
    P_a(\Omega) = -i\Omega +i\Delta_c+\frac k2,
    ~~~P_d(\Omega) = -i\Omega +i\Delta_s^0 + \frac \gamma 2
\end{equation}
If there were not coupling between the modes ($g_{eff}=0$) the cavity response to an external perturbation would be $1/P_a(\Omega)$.
It peaks when $\Omega = \Delta_c$ which is the cavity detuning from the pump.
The resonance width is {\rev $k$}.
Similarly,  $1/P_d(\Omega)$ describes the spin system response, with a resonance at $\Omega = \Delta_s$ and width $\gamma$.
The system then takes the form:

\begin{equation}
    \begin{cases}
        P_a(\Omega)\hat a(\Omega) =
        - ig_{eff}\hat d(\Omega) + \sqrt{k_1}\hat b_{in, 1}(\Omega) + \sqrt{k_2}\hat b_{loss}\\
        P_d(\Omega)\hat d(\Omega) = -i\xi\delta(\Omega) - ig_{eff}\hat a(\Omega) + \sqrt{\gamma}\hat{b}_f(\Omega)
    \end{cases}
\end{equation}
Solving for $\hat{a}(\Omega)$ we get:
\begin{equation}
    \hat{a}(\Omega) = \frac{1}{D(\Omega)} \left( P_d(\sqrt{k_1}\hat{b}_{in} + \sqrt{k_2}\hat{b}_{loss}) - g_{eff}(\xi\delta + i\sqrt{\gamma}\hat{b}_f \right),
\end{equation}
With denominator:
\begin{equation}
    D(\Omega) = P_{a}(\Omega)P_d(\Omega) + g_{eff}^2
\end{equation}
To describe the measurement process, we need to relate the field inside the resonator $\hat{a}(\Omega)$ to the field leaving the cavity through the input port.
This relation is given by the standard input-output relation which follows from the boundary conditions at the cavity walls:
\begin{equation}
    \hat b_{out}(\Omega) = \sqrt{k_1}\hat a(\Omega) - \hat b_{in}(\Omega)
\end{equation}
With expression for $\hat a(\Omega)$ we get the final expression:
\begin{equation}
    \hat b_{out}(\Omega) =
    \frac{k_1P_d-D}{D}\hat{b}_{in} + \frac{P_d\sqrt{k_1k_2}}{D}\hat{b}_{loss} - \frac{g_{eff}\sqrt{k_1}\xi}{D}\delta - \frac{ig_{eff}\sqrt{k_1\gamma}}{D}\hat{b}_f
\end{equation}
It is convenient to present this expression in a more compact form by introducing transfer functions for each noise source $K_{in}(\Omega), K_{loss}(\Omega), K_{f}(\Omega)$ and a useful signal $G(\Omega)$:

\begin{equation}
    G(\Omega) = -\frac{g_{eff}\sqrt{k_1}\xi}{D(\Omega)}
\end{equation}
\begin{equation}
    K_1(\Omega) \equiv K_{in}(\Omega) = \frac{k_1P_d(\Omega)}{D(\Omega)} - 1
\end{equation}
\begin{equation}
    K_2(\Omega) \equiv K_{loss}(\Omega) = \frac{P_d(\Omega)\sqrt{k_1k_2}}{D(\Omega)}
\end{equation}
\begin{equation}
    K_3(\Omega) \equiv K_{f}(\Omega) = - \frac{ig_{eff}\sqrt{k_1\gamma}}{D(\Omega)}
\end{equation}
Then the final expression for $\hat b_{out}(\Omega)$ is written as:
\begin{equation}\label{b_out_org}
    \hat b_{out}(\Omega) = G(\Omega)\delta(\Omega) + K_{in}(\Omega)\hat b_{in}(\Omega) + K_{loss}(\Omega)\hat b_{loss}(\Omega) + K_{f}(\Omega)\hat{b}_f
\end{equation}

\section{Homodyne Detection}\label{Appendix_homodyne_detection}
In this section, we describe in detail the transition from complex operators to quadrature components and the derivation of the homodyne signal.
Rewrite Eq.\,\eqref{b_out_org} as follows:
\begin{equation}
     \hat b_{out}(\Omega) = G(\Omega)\delta(\Omega)
     + K_i(\Omega) b_{i}(\Omega)
\end{equation}
where $i = in, out, loss$.
Using the quadrature amplitudes \eqref{b_c_b_s} and omitting the summation in $i$ for brevity, we obtain:
\begin{equation}
    \begin{cases}
        \hat b_{out}(\Omega) = G(\Omega)\delta(\Omega) + K_{i}(\Omega)\frac{b_{i}^c(\Omega) + ib_{i}^s(\Omega)}{\sqrt2}  \\

        \hat b_{out}^+(-\Omega) = G^*(-\Omega)\delta(\Omega) + K_{i}^*(-\Omega)\frac{b_{i}^c(\Omega) - ib_{i}^s(\Omega)}{\sqrt2}
    \end{cases}
\end{equation}

\begin{equation}
    \begin{cases}
        \hat b_{out}^c(\Omega) = G^c(\Omega)\delta(\Omega) + K_{i}^c(\Omega)b_i^c(\Omega) - K_i^s(\Omega)b_i^s(\Omega)  \\

        \hat b_{out}^s(\Omega) = G^s(\Omega)\delta(\Omega) + K_{i}^s(\Omega)b_i^c(\Omega) + K_i^c(\Omega)b_i^s(\Omega)
    \end{cases}
\end{equation}
The quadrature components for the signal transfer function $G(\Omega)$ and for an arbitrary complex transfer function $K(\Omega)$ are defined as:
\begin{equation}
    G^c(\Omega) = \frac{G(\Omega) + G^*(-\Omega)}{\sqrt2},~~~G^s(\Omega) = \frac{G(\Omega) - G^*(-\Omega)}{i\sqrt2}
\end{equation}
\begin{equation}
    K^c(\Omega) = \frac{K(\Omega) + K^*(-\Omega)}{2},~~~K^s(\Omega) = \frac{K(\Omega) - K^*(-\Omega)}{2i}
\end{equation}
The homodyne signal at local oscillator phase $\psi$ is the field projection onto the quadrature at angle $\psi$.
With homodyne detection, we can project the output field $ \hat b_{out}$ at any quadrature we choose, the choice is controlled by the local oscillator phase $\psi$.
The homodyne signal is proportional to operator:

\begin{align*}
    \hat b_{out}^{\psi}(\Omega) =& \hat b_{out}^c(\Omega)\cos{\psi} + \hat b_{out}^s(\Omega)\sin{\psi} =
    \\=& G_{\psi}(\Omega)\delta(\Omega) + (K^c_i(\Omega)\cos{\psi} + K^s_i(\Omega)\sin{\psi})b^c_i(\Omega) - (K^s_i(\Omega)\cos{\psi} - K^c_i(\Omega)\sin{\psi})b^s_i(\Omega) =
    \\=& G_{\psi}(\Omega)\delta(\Omega) + A_ib^c_i + B_ib^s_i,
\end{align*}
here:
$$
G_\psi(\Omega) = G^c(\Omega)\cos{\psi} + G^s(\Omega)\sin{\psi}
$$
This equation gives the field projection onto a quadrature rotated by $\psi$ from the amplitude quadrature.
By tuning the local oscillator phase $\psi$ we can measure the amplitude quadrature ($\psi=0$), the phase quadrature ($\psi=\pi/2$) or anything in between.
From this formula, we see that the homodyne signal contains the useful term $G_\psi(\Omega)\delta(\Omega)$ proportional to magnetic fluctuations, plus noise from each quadrature of each noise source, with weights $A_i, B_i$.
Since these coefficients depend on $\psi$ we can adjust the local oscillator phase to minimize the noise we do not want.

\section{Calculation of Spectral Densities}\label{Appendix_spectral_densities}
Here we explain how to compute the noise spectral densities for each noise source from the homodyne signal. The spectral density normalized to the signal $\delta(\Omega)$ is given by Eqs.\,\eqref{eq_correlate_function}, \eqref{eq_spectral_dens}. Note also that:
\begin{subequations}\label{S_sqz}
\begin{equation}
    S^{cc}_{in} = (n_{T}+\frac12)(ch~2r +sh~2r \cos{2\theta})
\end{equation}
\begin{equation}
    S^{ss}_{in} = (n_{T}+\frac12)(ch~2r - sh~2r \cos{2\theta})
\end{equation}
\begin{equation}
    S^{cs}_{in} = (n_{T}+\frac12)sh~2r \sin{2\theta}
\end{equation}
\end{subequations}
\begin{equation}
    S \equiv S^{cc}_{loss} = S^{ss}_{loss}=S^{cc}_{f} = S^{ss}_{f} = n_{th} + \frac12
\end{equation}
\begin{equation}
    S^{cs}_{loss} = S^{cs}_{f} = 0
\end{equation}
Therefore, the expression for the noise spectral density can be represented as the sum of three independent contributions:
\begin{equation}
    S_{\delta} = S_{\delta, in} + S_{\delta, loss} + S_{\delta, f},
\end{equation}
Here:
\begin{align*}
    S_{\delta, in} =& \frac{1}{|G_\psi|^2} \left( K^{cc}_{in}S^{cc}_{in} + K^{ss}_{in}S^{ss}_{in} + 2K^{cs}_{in}S^{cs}_{in} \right) =
    \\=& \frac{n_{T}+\frac12}{|G_\psi|^2} \left((K^{cc}_{in} + K^{ss}_{in})ch~2r + (K^{cc}_{in} - K^{ss}_{in})sh~2r\cos{2\theta} + 2K^{cs}_{in} sh~2r\sin{2\theta}\right)
\end{align*}
\begin{equation}
    S_{\delta, loss} =  \frac{n_{T}+\frac12}{|G_\psi|^2}(K_{loss}^{cc} + K_{loss}^{ss})
\end{equation}
\begin{equation}
    S_{\delta, f} =  \frac{n_{T}+\frac12}{|G_\psi|^2}(K_{f}^{cc} + K_{f}^{ss})
\end{equation}
Final equation:
\begin{equation}
    S_{\delta} = \frac{n_{T}+\frac12}{|G_\psi|^2}\left[(K^{cc}_{in} + K^{ss}_{in})ch~2r + (K^{cc}_{in} - K^{ss}_{in})sh~2r\cos{2\theta} + 2K^{cs}_{in} sh~2r\sin{2\theta} + K_{loss}^{cc} + K_{loss}^{ss} + K_{f}^{cc} + K_{f}^{ss}\right]
\end{equation}
To obtain the minimum spectral noise density it is necessary to optimize the expression for compression angle $\theta$ and homodyne detection angle $\psi$.
An important special case is case of $\Omega\rightarrow 0$ in which the minimum angle of $\theta$ has the analytical form:
\begin{equation}
    \min_{\theta} S_{\delta} = \frac{n_{th}+\frac12}{|G_\psi|^2}\left[(K^{cc}_{in} + K^{ss}_{in})ch~2r - sh~2r\sqrt{(K^{cc}_{in} - K^{ss}_{in})^2 + 4(K^{cs}_{in})^2} + K_{loss}^{cc} + K_{loss}^{ss} + K_{f}^{cc} + K_{f}^{ss}\right]
\end{equation}

\section{Coupling coefficient}\label{Appendix_coupling_coefficient}
To obtain reliable quantitative results it is important to choose the correct coupling strength between the quantum system and the cavity mode.
For a single photon, this coupling is given by the Rabi frequency $ {\rev g_1} = \Omega_R$.
For a large number of photons, the formula becomes:
\begin{equation}
    \Omega_R = {\rev g_1}\sqrt{N}
\end{equation}
Now we calculate the photon number N in the cavity.
At steady state, the input power  $P_{in}$ is equal to the loss power $P_{loss}$.
The loss power $P_{loss}$ is related to the energy stored in the cavity through the quality factor Q:
\begin{equation}
    P_{in} = P_{loss} = \frac{E\omega_c}{Q}, ~~~E=N\hbar\omega_c
\end{equation}
\begin{equation}\label{N_mw}
    N = \frac{QP_{in}}{\hbar\omega_c^2}
\end{equation}
For the given values, $\Omega_R = 2\pi\times 1~$MHz, $Q=20,000$, $P_{in}=0.01~$W, $\omega_c = 2\pi\times2790~$MHz, we get:
\begin{equation}
    N = 6.2\times 10^{15}
\end{equation}
\begin{equation}
    {\rev g_1} = \Omega_R/\sqrt{N} = 2\pi\times0.013~\text{Hz}
\end{equation}
The obtained value of the coupling coefficient ${\rev g_1}$ for one qubit can be generalized for a system of $N_q$ non-interacting qubits as follows:
\begin{equation}
    g_{eff} = {\rev g_1}\sqrt{N_{q}}
\end{equation}
{\rev The number of NV centers $N_q$ in the sample is determined by the NV concentration n (in ppm relative to the diamond lattice) and the sample volume V. Using the molar mass of diamond, we can write:
\begin{equation}
    N_{q} = \frac{\rho V}{M_r}N_An.
\end{equation}
Here, $\rho$ is the diamond mass density, $M_r$ is the molar mass of carbon, $N_A$ is Avogadro constant and n is the NV concentration in parts per million (ppm).
For a sample of dimensions 1x1x0.5 $mm^3$ and a concentration of n = 1ppm we calculate $N_{q}=9\times10^{13}$. Thus, for the ensemble $g_{eff} = 2\pi\times0.12$MHz.

From the expressions above, the effective coupling scales as $g_{eff} \propto \sqrt{nV}$.
This scaling has important consequences for magnetometer design.
Higher NV concentration increases $g_{eff}$ and thus improves the dispersive signal.
However, at high concentrations, dipole interactions between NV centers become significant, leading to additional dephasing, which degrades the sensitivity.
Larger sample volume increases $g_{eff}$ and the total number of sensing spins, improving sensitivity, but requires careful engineering of the microwave cavity mode to maintain spatial uniformity of the field.

For convenience, we have included all the fixed simulation parameters of the experiments in a table \ref{tab:sim_params}.

\begin{table}[htbp]
\centering
\caption{Fixed simulation parameters used throughout this work.}
\begin{tabular}{|c|c|c|}
\hline
\textbf{Parameter} & \textbf{Value} & \textbf{Description} \\
\hline
$\Delta_c$ & $0$ & Probe--cavity detuning \\
\hline
$\Delta_s$ & $2\pi \times 5$ MHz & Probe--spin detuning \\
\hline
$Q$ & $2 \times 10^4$ & Cavity quality factor \\
\hline
$k_1$, $k_2$ & $0.8k$, $0.2k$ & Input coupling and internal loss rates \\
\hline
$N_{NV}$ & $9 \times 10^{13}$ & Number of NV centers in the ensemble \\
\hline
$N$ & $6.2 \times 10^{15}$ & Mean number of photons in the cavity \\
\hline
$g_{eff}$ & $2\pi \times 0.12$ MHz & Collective NV--cavity coupling rate \\
\hline
$\gamma_e$ & $28$ GHz/T & Electron gyromagnetic ratio for NV \\
\hline
n & 1 ppm & Сoncentration of NV centers in the sample \\
\hline
\end{tabular}
\label{tab:sim_params}
\end{table}
}


\end{document}